\documentclass[12pt,preprint,useAMS,usenatbib]{emulateapj}
\usepackage{graphicx}
\usepackage{subfloat}
\usepackage{amsmath}
\usepackage{color}

\def\msun{\,{M_\odot}}
\def\Zsun{\,{Z_\odot}}
\def\mdot{\,{M_\odot {\rm yr^{-1}}}}

\def\spose#1{\hbox to 0pt{#1\hss}}
\def\lta{\mathrel{\spose{\lower 3pt\hbox{$\mathchar"218$}}
     \raise 2.0pt\hbox{$\mathchar"13C$}}}
\def\gta{\mathrel{\spose{\lower 3pt\hbox{$\mathchar"218$}}
     \raise 2.0pt\hbox{$\mathchar"13E$}}}

\newcommand{\etal}{{et al.\ }}
\def\HI{\hbox{H$\scriptstyle\rm I$}}

\def\HII{\hbox{H$\scriptstyle\rm II$}}

\def\CII{\hbox{C$\scriptstyle\rm II$}}

\def\kms{\,{\rm km\,s^{-1}}}

\def\cmm{\,{\rm cm^{-2}}}
\def\cm3{\,{\rm cm^{-3}}}

\lefthead{Shen et al. 2013}
\righthead{Baryon cycle of dwarf galaxies}
\submitted{accepted for publication in The Astrophysical Journal}


\begin{document}

\title{The Baryon Cycle of Dwarf Galaxies: Dark, Bursty, Gas-Rich Polluters}
\author{Sijing Shen$^1$, Piero Madau$^1$, Charlie Conroy$^1$, Fabio Governato$^2$, and Lucio Mayer$^3$} 
\altaffiltext{1}{Department of Astronomy and Astrophysics, University of California, 1156 High Street, Santa Cruz, CA 95064.}
\altaffiltext{2}{Astronomy Department, University of Washington, Seattle, WA 98195.}
\altaffiltext{3}{Institute of Theoretical Physics, University of Zurich, Winterthurerstrasse 190, CH-9057 Zurich, Switzerland.}

\begin{abstract}
We present results from a fully cosmological, very high-resolution, $\Lambda$CDM ``zoom-in" simulation of a group of seven field dwarf galaxies 
with present-day virial masses in the range $M_{\rm vir}=4.4\times 10^8-3.6\times 10^{10}\,\msun$. The simulation includes a blastwave scheme for supernova feedback, a star 
formation recipe based on a high gas density threshold, metal-dependent radiative cooling, a scheme for the turbulent diffusion of metals and thermal energy, and a uniform UV background
that modifies the ionization and excitation state of the gas. The properties of the simulated dwarfs are strongly modulated by the depth of the 
gravitational potential well. All three halos with $M_{\rm vir}<10^9\,\msun$ are devoid of stars, as they never reach the density threshold for star formation
of 100 atoms cm$^{-3}$. The other four, $M_{\rm vir}>10^9\,\msun$ dwarfs have blue colors, low star formation efficiencies, $-4.5\le \log M_*/M_{\rm vir}\le -2.5$,
high cold gas to stellar mass ratios, $0.2\le M_{\rm HI}/M_*\le 20$, and low stellar metallicities, $-2\le \langle {\rm [Fe/H]}\rangle \le -1$. 
Their bursty star formation histories are characterized by peak specific star formation rates in excess of $50-100$ Gyr$^{-1}$, far outside 
the realm of normal, more massive galaxies, and in agreement with observations of extreme emission-line starbursting dwarfs by the Cosmic Assembly Near-IR 
Deep Extragalactic Legacy Survey. The median stellar age of the simulated galaxies decreases with decreasing halo mass, with the two $M_{\rm vir}\simeq 2-3\times 
10^{9}\,\msun$ dwarfs being predominantly young, and the two more massive systems hosting intermediate and older populations. The two cosmologically young dwarfs are lit 
up by tidal interactions, have compact morphologies, and have metallicities and cold gas fractions similar to the relatively quiescent, extremely metal-deficient dwarf 
population that includes the recently-discovered Leo P. Metal-enriched galactic outflows produce 
sub-solar effective yields and pollute with heavy elements a Mpc-size region of the intergalactic medium, but are not sufficient to completely quench star formation 
activity and are absent in the faintest dwarfs. Within the limited size of the sample, our simulations appear to simultaneously reproduce the observed stellar mass and 
cold gas content, resolved star formation histories, and metallicities of field dwarfs in the Local Volume.  
\end{abstract}


\section{Introduction}\label{intro}
Dwarf galaxies are the smallest, most abundant, and least luminous systems in the Universe, and have come to play a critical
role in our understanding of how galaxies form and evolve. In the $\Lambda$CDM paradigm of cosmological structure formation, 
dwarfs collapsed early, become the building blocks for more massive objects, and hold clues to the reionization history of the 
Universe, the enrichment of the intergalactic medium (IGM), and the baryonic processes that shape the central dark matter density profiles of galaxies. 
Over the last two decades, observations of dwarf galaxies have challenged numerical simulations in $\Lambda$CDM and our understanding of the 
mapping from dark matter halos to their baryonic components. Two main classes of ``dwarf galaxy 
problems" have emerged both in the field and in the Galactic environment, and are currently faced by theoretical galaxy formation models. The first is an {\it abundance 
mismatch}. On the scale of dwarfs, the different shapes of the galaxy stellar mass function and the dark halo mass function require an efficiency of gas cooling 
and star formation in these systems that is more than 1 dex lower than that of Milky Way-sized halos \citep[e.g.,][]{Guo10,Behroozi13,Moster13}, a trend that that has been traditionally 
difficult to reproduce in cosmological hydrodynamical simulations \citep[see, e.g.,][and references therein]{Sawala11}. A strongly decreasing stellar mass fraction with 
decreasing halo mass is also required to solve the ``missing satellite problem", the discrepancy between the relatively small number of dwarf satellites orbiting the Milky Way 
(and M31) and the vastly larger number of dark matter subhalos predicted from $N$-body cosmological simulations \citep[e.g.,][]{Moore99,Klypin99,Diemand08,
Madau08,Koposov09,Rashkov12}. The second is a {\it structural mismatch}. Dark matter-only simulations predict steep (``cuspy") inner density profiles, but the 
observed rotation curves of dwarf galaxies show the sign of a near-constant density core \citep{Moore94,Flores94,deBlok02}. While in the field this long-standing issue has
become known as the ``core-cusp problem'', in the Galactic halo the same disagreement may be at the origin of the ``too-big-to-fail problem", whereby the most massive subhalos
found in dark matter-only simulations of Milky Way-sized systems are too dense to be consistent with existing constraints on luminous Galactic
satellites \citep{Boylan-Kolchin12}.

Emerging evidence suggests that a poor understanding of the baryonic processes involved in galaxy formation is at the origin of the dwarf galaxy puzzles. 
Photoheating from the cosmic ultraviolet background (UVB) after reionization has been suggested as a mechanism to suppress gas infall into small galaxy halos 
and set a minimum mass scale for galaxy formation \citep[e.g.,][]{Efstathiou92,Bullock00,Dijkstra04,Kravtsov04,Madau08}, but the only explanation   
for substantially modifying the dark matter distribution in field dwarfs that is consistent with $\Lambda$CDM relies on stellar feedback. Rapid mass loss 
driven by supernovae (SNe) has long been argued to reduce the baryonic content of luminous dwarfs \citep{Dekel86,Mori02,Governato07} and flatten their central 
dark matter cusps \citep{Read05}. A new generation of hydrodynamical simulations, with sufficient resolution to model clustered star formation in 
the highly inhomogeneous interstellar medium (ISM) and directly assess the impact of stellar feedback, reionization, and large-scale winds on the observable 
properties of galaxies, has been key in bringing the theoretical predictions in better agreement with many observations 
\citep[e.g.,][]{Mashchenko08,Governato10,Guedes11,Hopkins11,Governato12,Hopkins12,Zolotov12,Agertz13,Shen13,Stinson13,Teyssier13,Simpson13}.

In spite of the recent successes, the questions of whether realistic dwarf galaxies are a natural outcome of galaxy formation in $\Lambda$CDM remains open. 
Galactic outflows are observed to be ubiquitous in massive star-forming galaxies both at low and high redshifts \citep[e.g.,][]{Martin05,Veilleux05,Weiner09,Steidel10}, 
and an extended, patchy, metal-enriched circumgalactic medium (CGM) is detected in absorption around nearby late-type galaxies of all 
luminosities \citep[e.g.,][]{Tumlinson11,Stocke13}. Models of galaxy evolution require efficient outflows to explain the observed stellar masses, 
reduce the bulge-to-disk ratios, and account for the metals observed in the diffuse IGM, but their role in shaping the properties of 
dwarf galaxies is still poorly understood.
Large systematic surveys like the {\it Hubble Space Telescope} ACS Nearby Galaxy Survey Treasury (ANGST) program and the ALFALFA extragalactic 
\HI\ survey have recently provided uniformly measured star formation histories for nearby dwarfs \citep{Weisz11} and a large sample of very low \HI\ mass galaxies 
with complementary multiwavelength data \citep{Huang12}, enabling detailed investigations of the interplay between the gaseous and the time-resolved stellar components
in the faintest dwarfs. Studies of the metallicity distribution function and the relationship between the chemical abundances and the stellar content of dwarf satellites
are yielding new insights into their star formation efficiencies, infall of pristine gas, and outflows of enriched material \citep{Kirby08,Kirby11}. 
Observations of extreme emission-line galaxies by the {\it Hubble Space Telescope} Cosmic Assembly Near-IR Deep Extragalactic Legacy Survey (CANDELS) give strong 
indication that many or even most of the stars in present-day dwarf galaxies formed in strong, short-lived bursts at $z>1$ \citep{vanderWel11}.
Relatively little attention has been paid to comparing results from simulations with this avalanche of new data. Indeed, capturing the complex baryonic processes 
that regulate the ``metabolism" of dwarf galaxies over cosmic history requires cosmological hydro simulations of high dynamic range, something not easily achieved. 
Gas in such low-metallicity systems does not settle into a thin, dynamically cold disk \citep{Roychowdhury10,Leaman12}, and the low angular momentum support together 
with the shallow potential well make their ISM more susceptible to disruption from energetic SNe. As such, star formation may proceed in a ``bursty" manner that is 
different from that of larger mass spirals and that must be properly modelled. Stellar feedback may drive powerful, metal-enriched galactic outflows that modulate the stellar buildup, 
lower the gas fraction, and alter the chemical evolution of dwarfs. The strength of this effect depends on the ability to resolve a multi-phase ISM where star formation
and heating by SNe occur in a clustered fashion, on the depth of the potential well, and on the intensity of the star formation episode.

It is therefore important to evaluate a new set of $\Lambda$CDM high resolution simulations and test if galactic outflows are indeed key to forming realistic dwarf analogs.
In this work, we focus on the reduced baryonic content, high mass-to light ratios, star formation and metal enrichment histories 
of field dwarfs with $M_{\rm vir}\lta 10^{10.5}\,\msun$ using results from a fully cosmological $\Lambda$CDM simulation of large dynamic range. 
Our TreeSPH simulation -- one of the highest resolution ``zoom-ins" of field dwarfs run to redshift 0 -- 
includes a blastwave scheme for supernova feedback that generates large-scale galactic ouflows, a star formation recipe based on a gas density threshold 
comparable to the mean density of molecular cloud complexes, metal-dependent radiative cooling at all temperatures, and a scheme for the turbulent 
diffusion of thermal energy and metals. A spatially uniform UV background modifies the ionization and excitation state of the gas, photoionizing away abundant 
metal ions and reducing the cooling efficiency. As a validation of this approach, we will show that our simulations are able to simultaneously reproduce the main current 
observables in low-mass systems, from their stellar mass and cold gas content, through their resolved star formation histories, to their gas-phase and stellar 
metallicities. All such properties are strongly modulated by the depth of the gravitational potential well. 

\section{Simulation} \label{simulation}

The zoom-in simulation was performed with the parallel TreeSPH code \textsc{Gasoline} \citep{Wadsley04} in a $\Omega_M=0.24$, $\Omega_\Lambda=0.76$, $h=0.73$, 
$\sigma_{8}=0.77$, and $\Omega_b=0.042$ cosmology. The initial conditions were choosen to be the same as in the ``DG1" simulation of \citet{Governato10}.
The high resolution region, about 2 Mpc on a side at $z=0$, was embedded in a low-resolution dark matter-only periodic box of 25 Mpc on a side. It contains 
about 6 million dark matter and an equal number of SPH particles, with particle masses of  $m_{\rm DM}=1.6\times 10^4\,\msun$ and $m_{\rm SPH}=3.3\times 
10^3\,\msun$, respectively.  The gravitational spline softening length for collisional and collisionless particles was fixed to $\epsilon_G=86$ pc (physical) 
from $z=9$ to the present, and evolved as $1/(1+z)$ from $z=9$ to the starting redshift of $z=129$. In high density regions the gas smoothing length, $h$ 
(defined by the standard spline kernel, see \citealt{Monaghan92}), is allowed to shrink to $0.1\,\epsilon_G$, to ensure that hydrodynamic forces 
are well resolved on small scales.

\begin{figure*}
\centering
\includegraphics[width=0.7\textwidth]{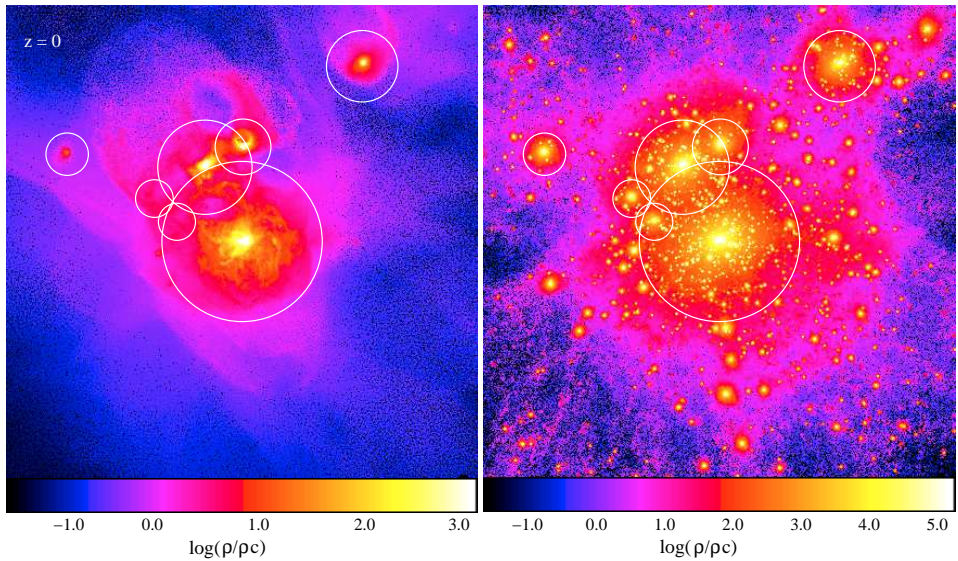}
\vspace{0.0cm}
\caption{Projected gas density ({left panel}) and dark matter density ({right panel}) of the simulated dwarf galaxy group at $z=0$, 
in a cube of 500 kpc on a side. The virial radii of the seven most massive halos are marked by the 
white circles. All densities are expressed in units of the critical density $\rho_c$. 
}
\label{fig1}
\end{figure*}

The star formation and blastwave feedback recipes are the same as in \citet{Governato10}. Briefly, star formation occurs stocastically when 
cold ($T<10^4$ K), virialized gas reaches a threshold density and is part of a converging flow, and follows a 
Schmidt law, 
\begin{equation}
{d\rho_*\over dt}=0.1 {\rho_{\rm gas}\over t_{\rm dyn}} \propto \rho_{\rm gas}^{1.5},
\label{eq:SFR}
\end{equation}
where $\rho_*$ and $\rho_{\rm gas}$ are the stellar and gas densities and $t_{\rm dyn}$ is the
local dynamical time. The large dynamic range of our simulation enables us to adopt a density threshold for star formation of 100 atoms cm$^{-3}$, 
as the local Jeans length at this density and a gas temperature of $T=10^3\,$K is resolved with more than 6 SPH smoothing lengths. 

\begin{table*}[t]
\centering
\caption{\enspace Present-Day Properties of the Simulated Dwarfs}\label{tab:7dwarfs}
\begin{tabular*}{\hsize}{@{\extracolsep{\fill}}lccccccccccc}
\\[-5pt]
\hline
\hline
\multicolumn{1}{l}{Name} &
\multicolumn{1}{c}{$M_{\rm vir}$} &
\multicolumn{1}{c}{$R_{\rm vir}$} &
\multicolumn{1}{c}{$V_{\rm max}$} &
\multicolumn{1}{c}{$V_{1/2}$} &
\multicolumn{1}{c}{$M_*$} &
\multicolumn{1}{c}{$M_{\rm gas}$} &
\multicolumn{1}{c}{$M_{\rm HI}$} &
\multicolumn{1}{c}{$f_b$} &
\multicolumn{1}{c}{$\langle$[Fe/H]$\rangle$} & 
\multicolumn{1}{c}{$M_V$} & 
\multicolumn{1}{c}{$B-V$}\\
\multicolumn{1}{l}{} &
\multicolumn{1}{c}{[$\msun$]} &
\multicolumn{1}{c}{[kpc]} &
\multicolumn{1}{c}{[$\kms$]} &
\multicolumn{1}{c}{[$\kms$]} &
\multicolumn{1}{c}{[$\msun$]} &
\multicolumn{1}{c}{[$\msun$]} &
\multicolumn{1}{c}{[$\msun$]} &
\multicolumn{1}{c}{} &
\multicolumn{1}{c}{} & 
\multicolumn{1}{c}{} \\
\hline
\\[-5pt]
Bashful & $3.59\times 10^{10}$ & $85.23$ & $50.7$ & $18.3$ & $1.15\times 10^8$ & $8.14\times 10^8$ & $2.34\times 10^7$ & $0.026$ & $-0.96\pm 0.51$ & $-15.5$ & 0.3\\
Doc     & $1.16\times 10^{10}$ & $50.52$ & $38.2$ & $21.6$ & $3.40\times 10^7$ & $1.74\times 10^8$ & $1.98\times 10^7$ & $0.018$ & $-1.14\pm 0.44$ & $-14.0$ & 0.4\\
Dopey   & $3.30\times 10^{9}$ &  $38.45$ & $22.9$ & $4.44$ & $9.60\times 10^4$ & $4.47\times 10^7$ & $1.96\times 10^6$ & $0.014$ & $-1.97\pm 0.44$ & $-8.61$ & 0.2\\
Grumpy  & $1.78\times 10^{9}$ &  $29.36$ & $22.2$ & $3.76$ & $5.30\times 10^5$ & $3.00\times 10^7$ & $5.40\times 10^5$ & $0.017$ & $-1.52\pm 0.54$ & $-11.0$ & 0.0\\
Happy   & $6.60\times 10^{8}$ &  $22.49$ & $15.6$ & \textemdash & \textemdash       & $2.54\times 10^6$ & \textemdash             & $0.004$ & \textemdash & \textemdash & \textemdash \\
Sleepy  & $4.45\times 10^{8}$ &  $19.71$ & $14.8$ & \textemdash & \textemdash       & \textemdash       & \textemdash & \textemdash & \textemdash & \textemdash & \textemdash \\
Sneezy  & $4.38\times 10^{8}$ &  $19.62$ & $13.2$ & \textemdash & \textemdash       & $1.64\times 10^5$ & \textemdash & $0.0004$ & \textemdash & \textemdash & \textemdash \\
\hline
\end{tabular*}
\tablecomments{Column 1 lists the dwarf names. The naming convention is based on the animated movie ``Snow White and the 
Seven Dwarfs". Columns 2, 3, 4, 5, 6, 7, 8, 9, 10, 11, and 12 give the present-day virial mass, virial radius (defined as the radius 
enclosing a mean density of 93 times the critical density), maximum circular velocity, circular velocity at the half-light radius, stellar mass, gas mass, 
\HI\ mass, baryon fraction $f_b\equiv (M_*+M_{\rm gas})/M_{\rm vir}$, mean stellar metallicity with dispersion, $V$-band magnitude, and $B-V$ color,  
respectively. The dark matter and initial gas and star particle masses of the simulation are $m_{\rm DM}=1.6\times 10^4\,\msun$, $m_{\rm SPH}=3.3\times 
10^3\,\msun$, and $m_*=10^3\,\msun$.
}
\vspace{0.cm}
\end{table*}

Each star particle has an initial mass of $m_*=10^3\,\msun$ and represents a simple stellar population with its own age, metallicity, and initial
mass function (IMF). Star particles inject energy, mass, and metals back into  the interstellar medium (ISM) through Type Ia and Type II SNe and stellar winds, 
following the prescriptions of \citet{Stinson06}. We track the formation of oxygen and iron separately, and convert oxygen to alpha-elements and iron to 
iron-peak elements assuming solar abundances patters \citep{Asplund09}. Each SN II deposits an energy of $10^{51}\,\epsilon_{\rm SN}\,$ergs 
into all the gas particles inside the blast radius, and the heated gas has its cooling shut off (to model the effect of feedback at unresolved scales) for the survival 
time of the hot, low-density cavity created by the remnant, $t_{\rm max}=(10^{6.85}\,{\rm yr})\, E_{51}^{0.32}n^{0.34}P_{-4}^{-0.7}$, where $n$ is the 
local ISM hydrogen density (in units of cm$^{-3}$), $P_{-4}$ is the ambient interstellar pressure (in units of $10^{-4}/k$), and $E_{51}$ is the 
total energy injected (in units of $10^{51}$ ergs) \citep{McKee77}. Cooling is turned off for all the particles within the blast radius, 
$R_E=(10^{1.74}\,{\rm pc})\,E_{51}^{0.32}n^{-0.16}P_{-4}^{-0.2}$ (for a maximum of 32 neighbors), and no kinetic energy is explicitly assigned to them. 
The energy injected by many SNe adds up to create larger hot bubbles and longer shutoff times than each individual supernova explosion. In combination with a high gas density 
threshold for star formation (which enables energy deposition by SNe within small volumes), this approach has been found to be key in producing realistic 
bulgeless dwarfs \citep{Governato10} and massive late-type spirals \citep{Guedes11}. 

The run analyzed here differs in a number of ways from the previously published DG1 simulation of \citet{Governato10}:
1) the IMF follows the modern determination by \citet{Kroupa01}. This increases the number of Type II supernovae per unit stellar mass 
by about a factor of two, and the IMF-averaged metal yield by a factor of three compared to the \citet{Kroupa93} IMF used by \citet{Governato10}; 2) the 
fraction of SN energy that couples to the ISM is $\epsilon_{\rm SN}=1$ as in \citet{Governato12}, but 2.5 times higher than in \citet{Governato10}; 
3) metallicity-dependent 
radiative cooling is included at all temperatures in the range 100-$10^9$ K. The cooling function is determined using pre-computed tabulated
rates from the photoionization code \textsc{Cloudy} \citep{Ferland98}, following \citet{Shen10}. \textsc{Cloudy} tables assume that metals are in 
ionization equilibrium. The ionization, cooling, and heating rates for primordial species (H, H$^+$, He, He$^+$, He$^{++}$) are calculated time-dependently 
from the rate equations; 
4) a uniform UVB  modifies the ionization and excitation state of the gas, photoionizing away abundant metal ions and reducing the cooling efficiency.
It is implemented using the new \citet{Haardt12} redshift-dependent spectra, including emission from quasars and star-forming galaxies. As in previous runs, the
gas is assumed to be optically thin to ionizing radiation at all wavelengths; and 5) a scheme for turbulent mixing in shearing flows that redistributes heavy elements 
and thermal energy between wind material and the ambient gas is included following \citet{Shen10,Shen13}. 

\begin{figure*}
\centering
\includegraphics[width=0.495\textwidth]{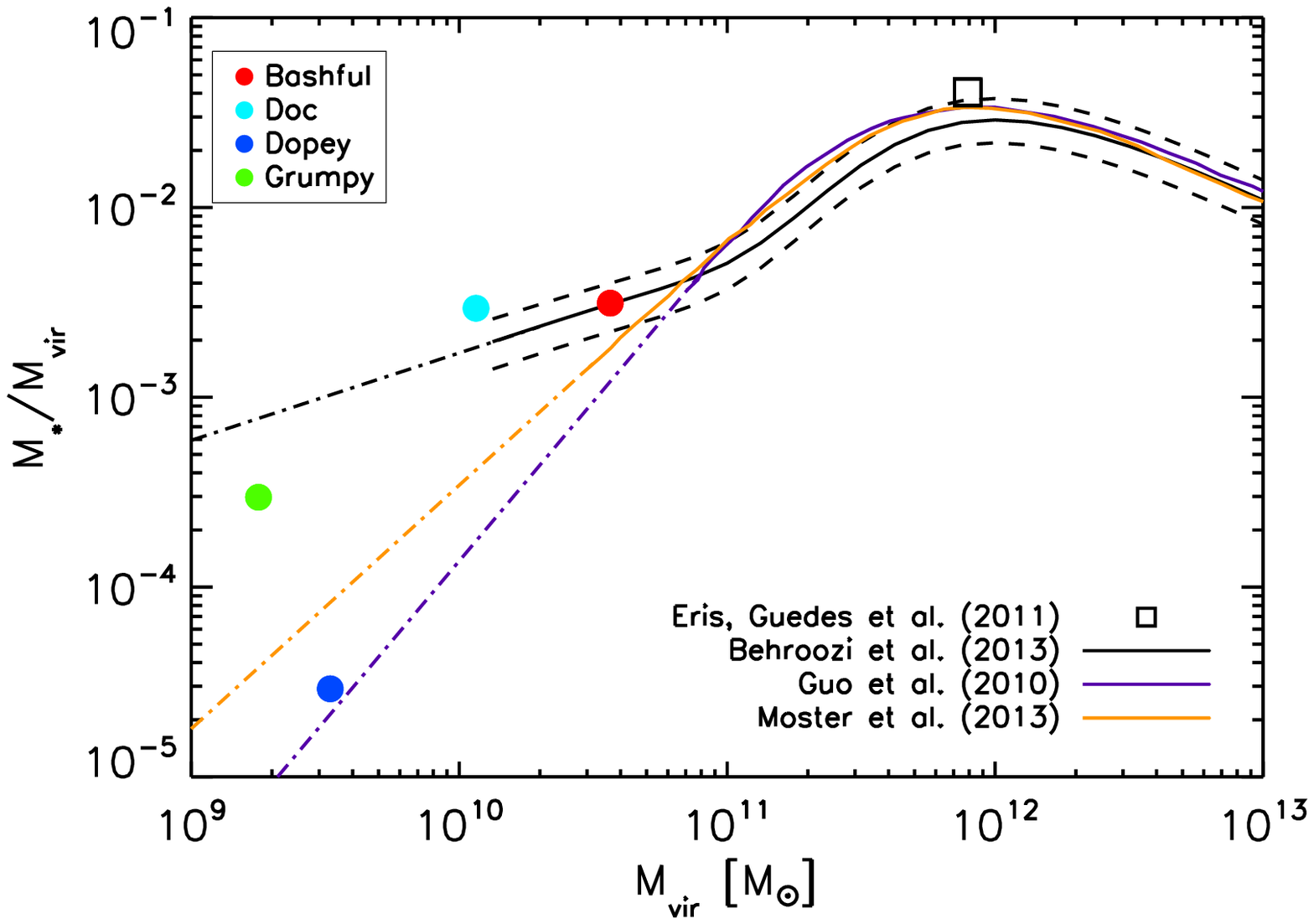}
\includegraphics[width=0.495\textwidth]{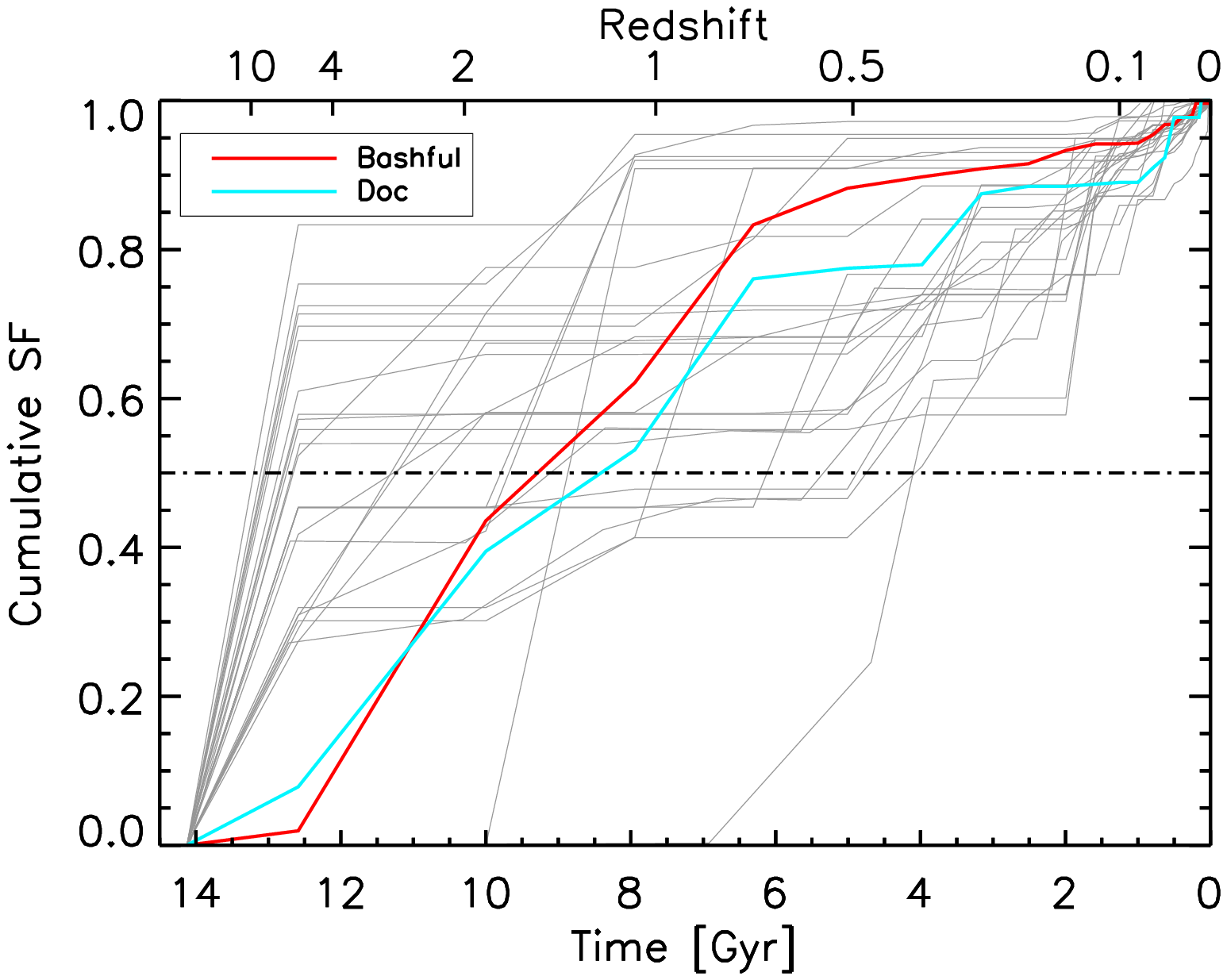}
\caption{{Left panel:} The stellar mass fraction of dwarf galaxies at $z=0$. {Filled colored circles:} the simulated ``Bashful", ``Doc", ``Dopey", and ``Grumpy" dwarfs.
{Empty square:} the late-type Milky Way analog ``Eris" \citep{Guedes11}, run using the same star formation and supernova feedback recipes but without metal-line cooling at $T > 10^{4}$ K.  
{Black, orange and purple solid lines:} the mean present-day stellar mass-halo 
mass relations of \citet{Behroozi13}, \citet{Moster13} and \citet{Guo10}, respectively, with the extrapolations below a few times $10^{10}\,\msun$ indicated with 
the dot-dashed lines in the corresponding colors. The black dashed lines show the $\pm 1\sigma$ limits of the SMHM relation of \citet{Behroozi13}.
{Right panel:} Cumulative star formation 
history, i.e. the fraction of total stellar mass formed prior to a given redshift or lookback time. {Grey lines:} 
individual dwarf irregulars in the ANGST sample \citep{Weisz11}. The horizontal dot-dashed line represents 50\% of the total stellar mass today.  
{Thick colored lines}: ``Bashful" and ``Doc". The simulation results have been binned as the ANGST data for better comparison. 
}  
\vspace{+0.cm}
\label{fig2}
\end{figure*}

We do not attempt to follow the atomic-to-molecular transition. 
Simulations of dwarf galaxies with H$_2$-regulated star formation and comparable resolution as achieved here have been recently run to $z=0$ by 
\citet{Christensen12} and \citet{Governato12} using a time-dependent H$_2$ chemical network, or to high redshift by \citet{Kuhlen12} using the two-phase equilibrium 
approximation for the \HI\ to H$_2$ transition of \citet{Krumholz09}. While molecular hydrogen is not an important coolant in present-day galaxies (CO and \CII\ are), 
it is an excellent proxy for the presence of cold gas. Linking star formation to the local H$_2$ fraction allows the adoption of a well motivated and
robust star formation prescription that is based on the observed universal efficiency of gas conversion per free-fall time of molecular clouds in the disks
of spiral galaxies \citep{Bigiel11}. However, non-equilibrium self-consistent chemical networks are complicated and expensive to implement in cosmological 
simulations, as the effective H$_2$ formation rate depends on the gas clumping factor on unresolved scales, the intensity of the photodissociating 
interstellar radiation field, the self-shielding and the shielding by dust grains \citep{Gnedin09}. The threshold we choose, 100 atoms cm$^{-3}$, is close to the volume density 
at which atomic-to-molecular transition occurs in simulations of \citet{Christensen12} for low metallicity gas. Moreover, a strong association between star formation and H$_2$ 
and a lack of association with \HI\ are only found down to metallicities of about 1/5 of solar, the lowest metallicity systems measured \citep{Bolatto11}. 
At the low gas metallicities of our simulated dwarfs, $Z/Z_\odot=0.13-0.013$, there are no observational calibrations of the molecular gas content.  
Theoretical models also suggest that, at metallicities below a few percent of the solar value, the correlation between H$_2$ and star formation breaks 
down and stars form in the cold atomic phase \citep{Krumholz12}. Cosmological simulations of dwarf galaxy formation that incorporate H$_2$-regulated star formation 
and a weak form of stellar supernova feedback have been presented elsewhere \citep{Kuhlen13}.   

\begin{figure*}
\centering
\includegraphics[width=0.495\textwidth]{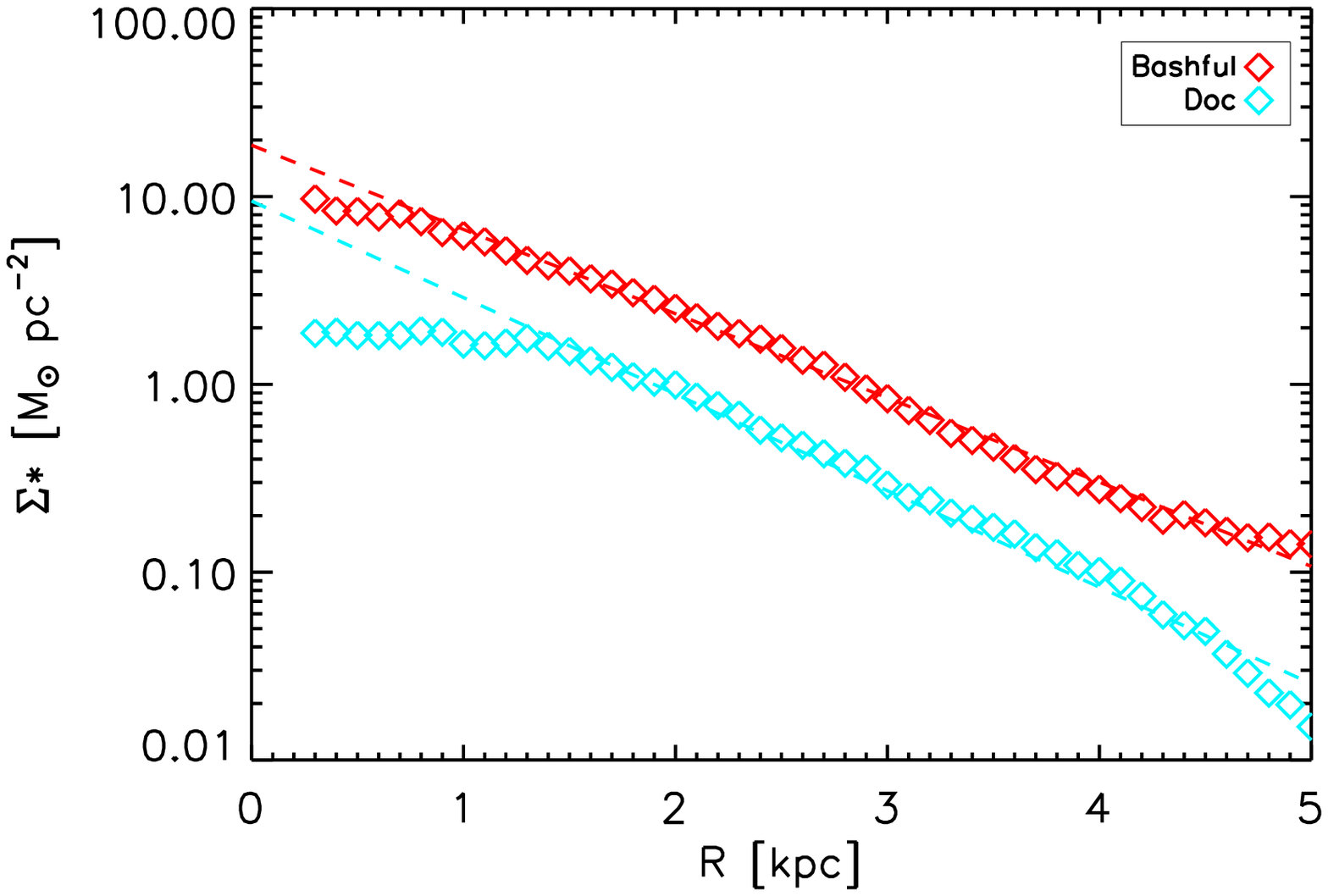}
\includegraphics[width=0.495\textwidth]{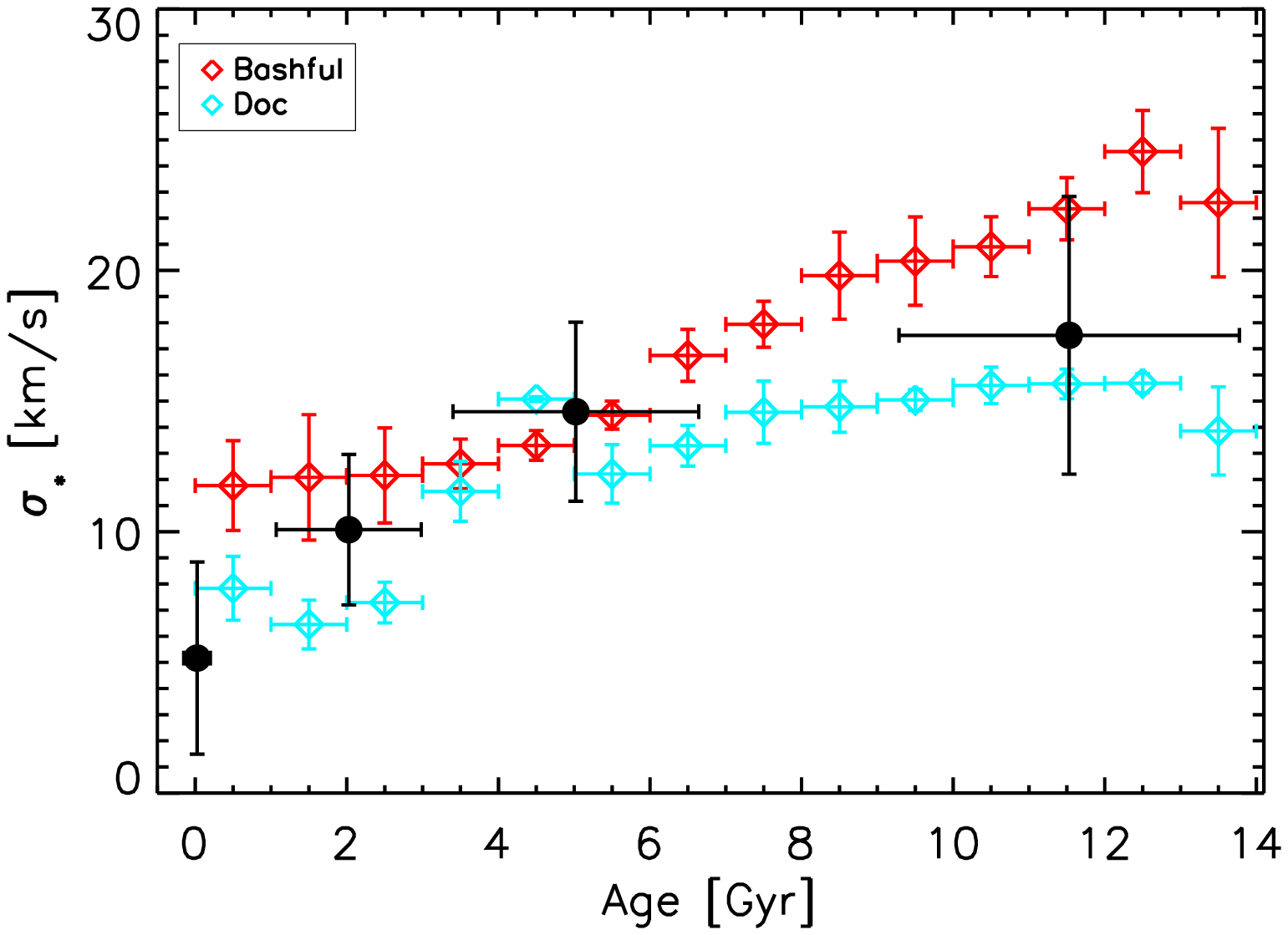}
\caption{{Left panel:} The 1D line of sight stellar surface density profiles of Bashful and Doc. The dashed lines show exponential fits to their stellar disks,
$I(R>1\,{\rm kpc})=I_d\exp(-R/R_d)$, with scale lengths $R_d=0.97$ kpc for Bashful, and $R_d=0.84$ kpc for Doc. The two dwarfs are bulgeless, 
with Doc showing the clear signature of a stellar core within the central 1 kpc.
{Right panel:} The present-day line of sight velocity dispersion as a function of age in the central ($<2$ kpc) stellar components of Bashful and Doc. 
Younger stellar cohorts are kinematically colder than older cohorts. 
{Filled black dots:} observations of the isolated dwarf irregular galaxy WLM \citep{Leaman12}. 
} 
\vspace{+0.cm}
\label{fig3}
\end{figure*}

\section{Properties of the Simulated Galaxies}

The selected galaxy forming region is depicted in Figure \ref{fig1}. The image shows projected gas and dark matter densities at $z=0$ 
in a cube 500 kpc on a side. The virial radii of the seven, well-resolved, most massive members of what appears to be a small dwarf galaxy 
group are marked by the white circles in the top panel.  The present-day properties of the seven dwarfs are summarized in Table \ref{tab:7dwarfs}.
Their virial masses today are in the range $4.4\times 10^8-3.6\times 10^{10}\,\msun$, with the most massive system being resolved by more than 
2.2 million dark matter particles, and the lightest one by just 27,000. The maximum circular velocities bracket the interval $13-51\,\kms$.  
The two most massive dwarfs are within 100 kpc of each other and form an isolated  dwarf-dwarf galaxy pair like those found in the {Sloan Digital 
Sky Survey} by \citet{Geha12}. Three smaller dwarf halos are in the process of merging with the pair. All are ``field" dwarfs with the nearest 
massive halo ($M_{\rm vir}=2.5\times 10^{12}\,\msun$) more than 3 Mpc away. 

\subsection{Two Luminous Dwarfs}

Stellar masses are strongly modulated by the depth of the gravitational potential. The two, $M_{\rm vir}\gta 10^{10}\,\msun$ most massive 
dwarfs, ``Bashful" and ``Doc", have stellar masses of $M_*=11.5\times 10^7\,\msun$ and $3.4\times 10^7\,\msun$, respectively. Both are gas-rich, show knots 
of recent star formation, have blue colors, and would be classified as late-type dwarf irregulars (dIrrs). We have used the Flexible Stellar Population Synthesis models of 
\citet{Conroy09} to generate broadband luminosities of composite stellar populations of different ages and metallicities.
Bashful and Doc have present-day $V$-band magnitudes of $M_V=-15.5$ and $-14$, respectively. Their stellar mass fractions, 
$M_*/M_{\rm vir}\simeq 0.003$, are in excellent agreement with the stellar mass-halo mass (SMHM) relation recently derived for present-day dwarfs by \citet{Behroozi13}, 
although Doc lies above the power-law extrapolations of the SMHM relations of  \citet{Moster13} and \citet{Guo10} (see Figure \ref{fig2}). \citet{Behroozi13} shows that the low mass 
power-law behavior of the SMHM relation is broken on dwarf galaxy scales down to $10^{10} \msun$, corresponding to an upturn in the stellar mass function below $10^{8.5}\,\msun$ \citep{Baldry08}. 
Although the exact behaviour of the mean SMHM relation below a few times $10^{10} \msun$ is still under debate \citep{GK14,Brook14}, we note that the scatter of the SMHM relation for very low mass 
dwarf galaxies is also significant \citep{Kuhlen13}. We have made no attempt to correct the true stellar masses directly measured in the simulation by the systematic photometric bias discussed 
in \citet{Munshi13} (see also \citealt{Guedes11}), which can lead to stellar mass errors of up to 50\% for individual galaxies. 
Note also how, as a consequence of the larger injection of energy by SNe per unit stellar mass assumed here, Bashful forms four times less stars than its 
DG1 counterpart in the \citet{Governato10} simulations, a fact that underscores the sensitivity of the predicted star formation efficiency to the strength 
of stellar feedback. 

\begin{figure*}
\centering
\includegraphics[width=0.7\textwidth]{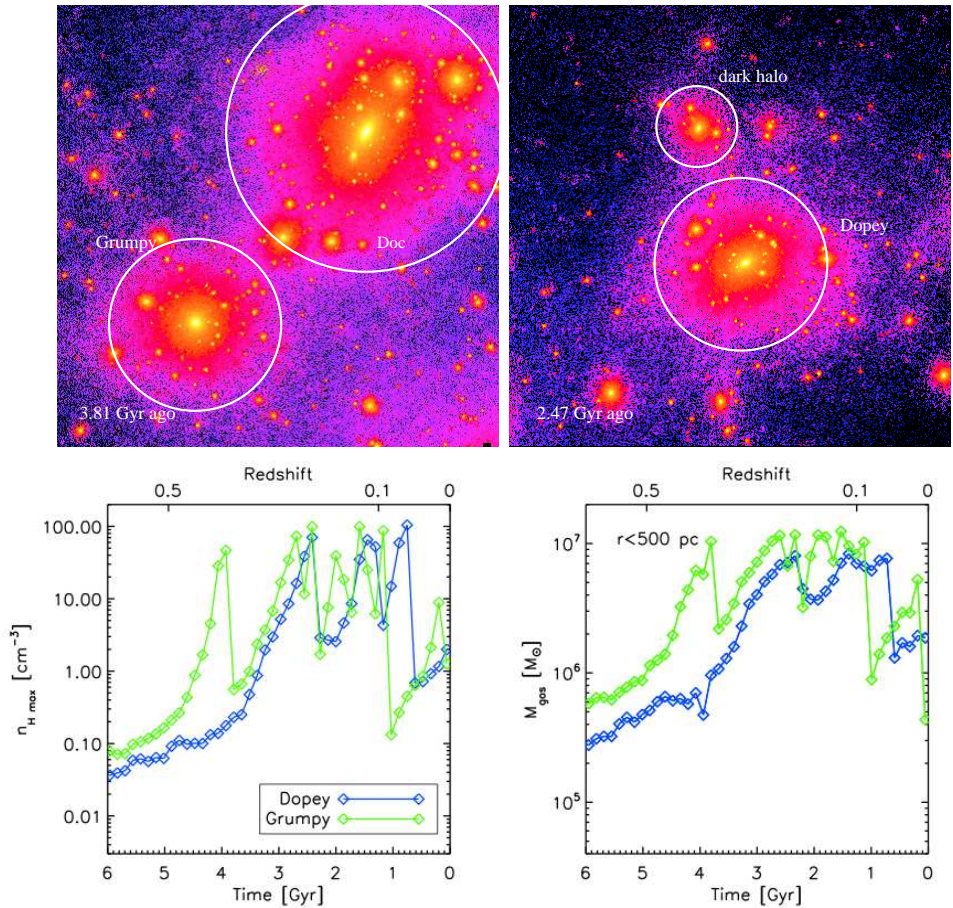}
\caption{The late star formation cycle of the two faint, extremely metal-deficient simulated dwarfs Dopey and Grumpy. {Top panels:} Projected 
dark matter density of the simulated dwarf galaxy group at lookback times 2.47 Gyr ({right}) and 3.81 Gyr ({left}), when Dopey and Grumpy undergo 
their first burst of star formation. In both panels the box size is 780 comoving kpc, the color scale is the same of Figure \ref{fig1}, 
and virial radii are marked by the white circles. The physical distance between Grumpy and Doc in the left panel is 90 kpc.   
{Lower panels:} The rise in the maximum gas density ({left}) and central gas mass ({right}) that precedes 
the first (and subsequent) starbursts. The gas finally reaches the threshold of 100 atoms cm$^{-3}$ and a cycle of late star formation is initiated where
each burst is followed by a reduction in the total gas content of the central regions. 
}
\vspace{+0.cm}
\label{figtidal}
\end{figure*}

Resolved stellar populations have proven to be a very powerful tool for observationally constraining scenarios of dwarf galaxy evolution, as past patterns 
of star formation and chemical evolution are encoded in a galaxy's optical color-magnitude diagram. Figure \ref{fig2} compares the cumulative star formation histories (SFHs), 
i.e. the fraction of total stellar mass formed prior to a given cosmic time, of individual dIrrs in the ANGST sample \citep{Weisz11} with those of our two most luminous 
simulated dwarfs. While the SFHs of individual ANGST galaxies are quite diverse, most dwarfs are not entirely old stellar populations, i.e. they have intermediate or 
recent star formation. The {\it average} dwarf formed the bulk of its stars prior to redshift 1, exhibits dominant ancient star formation ($>10$ Gyr ago) and  lower levels of activity 
over the last 6 Gyr, and produced about 8\% of its total stellar mass within the last 1 Gyr. The cumulative SFHs of Bashful and Doc appear to be broadly consistent with the observations, 
as both formed about half of their stars more than 9-10 Gyr ago, 85\% and 76\% of them prior to 6 Gyr ago ($z\gta 0.7$), and 6\% and 11\% over 
the last 1 Gyr, respectively. We find no evidence on these halo mass scales for the need of early stellar feedback (in addition to SN feedback) 
in order to match the evolving SMHM relation and curtail the overproduction of stars at high redshift \citep{Stinson13,Aumer13,Moster13}.
While uncertainties in the recovered SFHs of ANGST galaxies at the earliest bin are significant, due to the shallow nature of the data and accuracies of
the models of evolved stars, the simulation may in fact produce too few stars at $z>4$, something we will explore further in future work. 

At the present epoch, both Bashful and Doc have exponential stellar disks of scale lengths $R_d=0.97$ kpc and $R_d=0.84$ kpc, respectively, extending to 5 kpc 
(see the left panel of Figure \ref{fig3}), with no bulge in the center. Doc shows the clear signature of a stellar core within the central 1 kpc,  a size that is similar 
to the stellar core in the isolated dIrr galaxy WLM \citep{Leaman12}. In Figure \ref{fig3} we also show (right panel) the velocity dispersion of stellar populations of different 
ages inside a radius of 2 kpc, where the velocity distribution is close to Gaussian. There is a trend in both systems of an increasing velocity dispersion with increasing stellar age. Stars that 
formed early are born in a compact configuration and quickly scattered into a more extended, kinematically hot component by galaxy mergers and outflows, while subsequent cohorts 
form in progressively colder configurations from gas with increasing levels of rotational support (see also \citet{Brook04, Bird13}. A similar trend has been recently observed in WLM, 
and the data points from \citet{Leaman12} are also plotted in the figure. With a dynamical mass of about $\sim 10^{10}\,\msun$, a stellar mass of $10^7\,\msun$, and an \HI\ mass of 
$6\times 10^7\,\msun$ \citep{Leaman12}, WLM closely resembles our simulated Doc. As in WLM and in the isolated dwarf simulated with a strong feedback scheme 
by \citet{Teyssier13}, both Bashful and Doc are characterized by a dynamically hot stellar structure. 

\begin{figure*}
\centering
\includegraphics[width=0.8\textwidth]{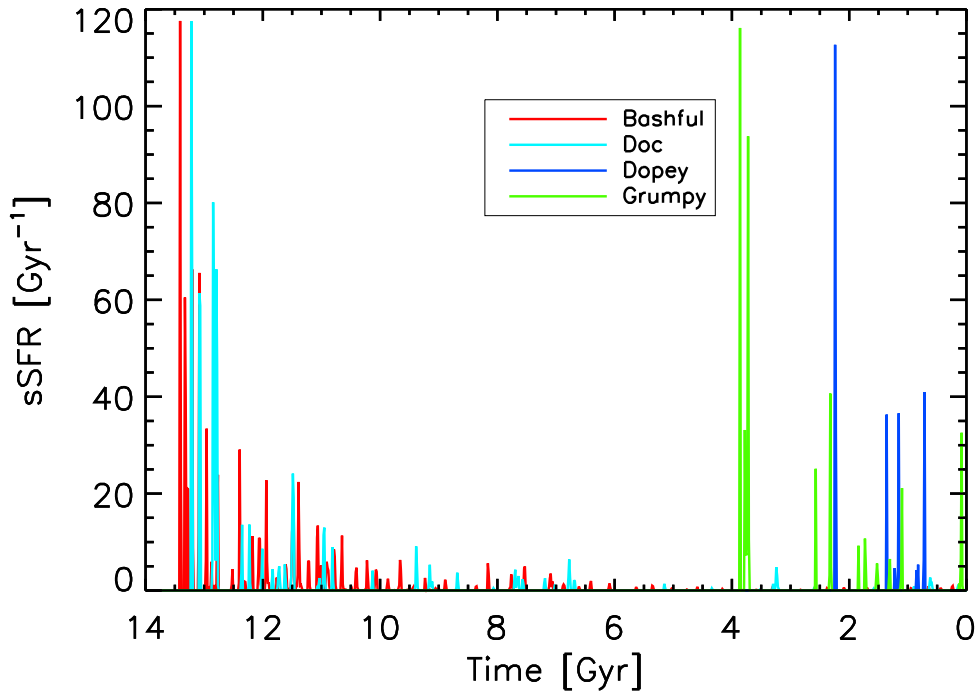}
\caption{The bursty star formation histories of the simulated dwarfs. The specific star formation rate (sSFR) is plotted againt lookback time. 
}
\vspace{+0.cm}
\label{fig4}
\end{figure*}

\subsection{Two Faint Dwarfs}

The two intermediate-mass dwarfs ``Dopey" and ``Grumpy" (see Table \ref{tab:7dwarfs}) fall well below the extrapolation of the SMHM relation below 
$M_{\rm vir}=10^{10}\,\msun$ of \citet{Behroozi13}. Grumpy lies above the power-law extrapolations of the SMHM relations of \citet{Guo10} and \citet{Moster13}, 
while Dopey appears to be consistent with these two extrapolations. Dopey and Grumpy have virial masses of about $2-3\times 10^9\,\msun$, baryon fractions above 1\% 
(comparable to Bashful and Doc), and maximum circular velocities in excess of $20\,\kms$, yet they manage to form by $z=0$ only $10^5$ and $5\times 10^5\,\msun$ of 
stars, i.e. their star formation efficiencies are $2\times 10^{-5}$ and $3\times 10^{-4}$, respectively. At the present epoch, Dopey and Grumpy are very faint, 
with $M_V=-8.6$ and $-11$, respectively, and extremely metal poor (see below). Both are cosmologically young as they started forming stars only 
3-4 Gyr ago, and have very blue colors, $B-V=0.2$ (Dopey) and 0.0 (Grumpy). They have very compact morphologies with half-light radii of 80-90 pc. 
The blue color and compactness are similar to the properties of the extremely metal-deficient blue compact dwarfs (XBCD, \citealt{Papaderos08}). As it has been argued 
to be the case for the prototypical XBCD I Zw 18, the older stellar populations (with age $>$ 1 Gyr) in Dopey and Grumpy have not settled into an extended, low surface brightness envelope.  
However, an important difference between our two faint dwarfs and XBCDs is the central surface brightness. Whereas XBCDs have nuclei
with visual central surface brightness in the range 20-23 mag arcsec$^{-2}$, Dopey and Grumpy have surface brightness of, respectively, 24.0 and 26.5 in the B-band. As such, they are more 
similar to low surface brightness dwarfs in the Local Group \citep[e.g.][]{Mateo98,Weisz11}. Moreover, contrary to the starbursting XBCD I Zw 18 \citep{Skillman93,Izotov99}, Dopey and 
Grumpy appear to lie on the gaseous mass-metallicity relation (see Section \ref{metals}). Among local dIrrs with deep HST photometry, the closest analog to Dopey and Grumpy in terms 
of star formation history is Leo A \citep{Cole07}. Leo A is also the only local dIrr that formed as much as 90\% of its stars in the last 8 Gyr, 
well after the reionization epoch \citep{Skillman13}. With virial temperatures close to the temperature of the photoionized IGM, it is only at late epochs that gas in these dwarfs 
can reach our threshold density for star formation. It is tempting to argue that this is exactly the kind of evolutionary history that Leo A underwent.
 
\begin{figure}
\centering
\includegraphics[width=0.495\textwidth]{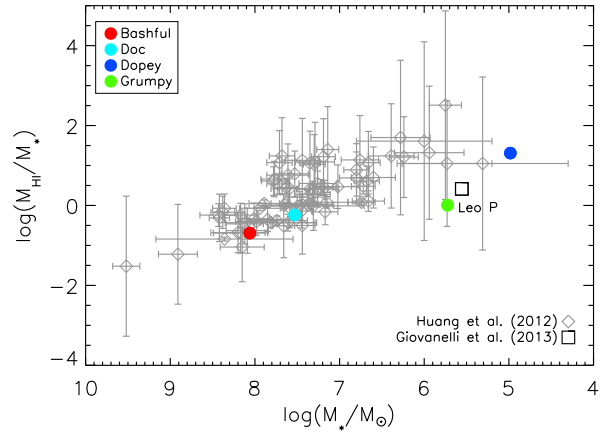}
\caption{Cold gas fraction, $M_{\rm HI}/M_*$, and star formation. Low stellar-mass galaxies in the ALFALFA dwarf sample ({empty gray diamonds}) are 
more \HI\ gas rich \citep{Huang12}. {Empty large square:} Leo P \citep{Giovanelli13}. {Colored solid dots}: the simulated dwarfs. Note that Grumpy 
gets stripped of 60\% of its gas during the interaction with the more massive Doc, and therefore has a relatively low present-day cold gas fraction.}  
\vspace{+0.cm}
\label{figfHI_Ms}
\end{figure}

Despite all the differences, it is intriguing to compare the origins of Dopey and Grumpy to that of the XBCDs. The origin of XBCDs has been ascribed to 
tidal interactions or merging between gas-rich dwarfs having undergone little previous star formation and enrichment, and possibly co-evolving in a group of low-mass 
systems. This is indeed the formation path of Dopey and Grumpy. The top panel of Figure \ref{figtidal} shows an image of the dark matter density field 3.81 Gyr 
and 2.47 Gyr ago, at the time of their first burst of star formation, when Grumpy is being tidally perturbed by Doc (left image) and Dopey is merging with a smaller dark halo 
(right image). One consequence of such interactions is the rapid inflow of baryons within the inner 500 pc (lower right panel) that fuels the central starburst. 
The gas can finally reach the threshold of 100 atoms cm$^{-3}$ (lower left panel), and a cycle of bursty late star formation activity is initiated. 
During the interaction with the more massive Doc, Grumpy gets stripped of 60\% of its gas, something that is reflected in its relatively low present-day cold gas fraction
(see Figure \ref{figfHI_Ms}). We note here that the lack of a stellar population with age comparable to the Hubble time may be attributed to our implementation of reionization, as the UVB
is switched on instantaneously and gas self-shielding is not modeled in our simulations. To gauge the impact of the UVB, we have performed a twin simulation with the same star formation 
and feedback recipes, but no UVB. Even in this case, neither Dopey nor Grumpy form any stars before redshift 6, i.e. before the epoch when reionization is known to be essentially completed. 
Our simulations do not include H$_{2}$ cooling, however, an approximation that may prevent gas to condense and form stars before reionization \citet{Christensen12}. 
We defer a more detailed study of the impact of self-shielding and reionization on the star formation histories of dwarfs to a future paper.

The median stellar ages of Dopey and Grumpy are only about 1.1 Gyr, compared to the 9.7 Gyr and 9.1 Gyr of Bashful and Doc. Within our limited sample, the median stellar age
decreases with decreasing halo mass, with galaxies with total masses below $10^{10}\,\msun$ being predominantly young, and more massive systems hosting intermediate and 
older populations. 


\subsection{Bursty Star Formation Histories}

The star formation histories of the simulated dwarfs, binned over a time-interval of 14 Myr, are shown in Figure \ref{fig4}. As in previous dwarf
galaxy simulations with effective SN feedback \citep[e.g.,][]{Governato10,Teyssier13}, the star formation activity is extended, stochastic, with large amplitude bursts followed by
short quiescent phases. All the four dwarfs show peak specific star formation rates, ${\rm sSFR}\equiv \dot M_*/M_*$, in excess of $50-100$ Gyr$^{-1}$, far outside
the realm of normal, more massive galaxies.  The amount of energy deposited into the ISM by SN explosion during the strongest bursts in Bashful and Doc can approach $10^{56}$ erg.  
Each star formation episode is typically preceded by an increase in the central gas supply, and is accompanied by a reduction in the total baryonic mass as SN-driven outflows 
deplete the central regions of star-forming material. A new cycle starts again as fresh gas cools and is reaccreted from the halo, sinks to the center of the potential well, and 
triggers another starburst. It is the potential fluctuations generated by these cycles of gas inflows and rapid outflows following centrally-concentrated bursts of star 
formation that irreversibly transfer energy into collisionless particles and generate dark matter cores according to \citet{Pontzen12}. 

Bursts of star formation have long been invoked to explain the observed properties and colors of faint blue dIrrs \citep{Searle73}.
Resolved observations of stellar populations in the nearest low-mass systems show that the temporal separation of bursts of star formation ranges from tens or hundreds of Myr in
dIrrs to Gyr in dwarf spheroidals (dSphs) \citep[see, e.g.,][]{Tolstoy09}. No long interruptions in the star formation activity are seen in our simulations. Both Bashful and
Doc have undergone episodes of star formation within the past 150 Myr, but only Bashful is actively forming stars today. A late bursty SFH may provide an explanation for
the systematic discrepancy observed in star-forming dwarfs between star formation rates inferred from their FUV continua and H$\alpha$ nebular emission \citep{Lee09},
as the latter is emitted on the timescale of O-star lifetimes ($\lta 5$ Myr). Similarly, in the ALFALFA dwarf sample, FUV magnitudes appear to overpredict star formation rates 
below $10^{-2}\,\mdot$ compared to estimates derived from SED fitting, another evidence for the bursty nature of star formation in dwarfs \citep{Huang12}. 
An abundant population of extreme emission-line galaxies has been identified at $z\sim 1.7$ by the Cosmic Assembly Near-IR
Deep Extragalactic Legacy Survey (CANDELS). With stellar masses $M_*\sim 10^8\,\msun$ and [OIII] emission lines with rest-frame equivalent width $\sim 1000$ \AA, these 
starbursting dwarfs are  growing at rates (sSFR) of $20-200$ Gyr$^{-1}$ \citep{vanderWel11}, in agreement with  our simulated systems.
These observations provide a strong indication that many or even most of the stars in present-day dwarf galaxies formed in strong, short-lived bursts at $z>1$.

We finally note that both Bashful and Doc have 4000 \AA\ breaks, $D_n4000$, that are below 1.4, i.e. they would be classified as star-forming according to a $D_n4000$ cut. 
This is in agreement with the results of \citet{Geha12}, who find that dwarf galaxies with $10^7<M_*<10^9\,\msun$ and no active star formation are extremely rare in the 
field. In more dense environments, quenched galaxies account for 23\% of the dwarf population over the same stellar mass range.

\subsection{Cold Gas Properties}

Cold gas is the fuel needed to sustain star formation, and while star formation is observed to be more directly linked 
to the molecular interstellar component at metallicities above few tens of solar, theoretical models predict stars to form in 
the cold atomic phase at extremely low metallicities \citep{Krumholz12}. Furthermore, dwarf galaxies appear to be significantly 
fainter in CO than a simple linear scaling with galaxy mass would suggest \citep{Schruba12}. The ALFALFA extragalactic \HI\ survey has recently  
provided a large sample of very low \HI\ mass galaxies with complementary multiwavelength data from the {Sloan Digital Sky Survey}
(SDSS) and {\it Galaxy Evolution Explorer} (GALEX), and enabled statistical studies of the interplay between the gaseous and stellar components
in the faintest dwarfs \citep{Huang12}. In Figure \ref{figfHI_Ms}, we compare the cold gas fraction, $M_{\rm HI}/M_*$ in our simulated dwarfs
with the {\it s-sed} galaxies of \citet{Huang12} with reliable UV/optical magnitudes (74 of them). The neutral hydrogen mass in each galaxy is computed directly from the simulation, 
where the \HI\ fractions of gas particles are evolved by integrating the ionization equations for primordial ion species. More than half (46) of the {\it s-sed} dwarfs have $M_{\rm HI}/M_*>1$, 
meaning that their baryonic mass is dominated by atomic gas, rather than by stars: the mean \HI\ mass grows with $M_*$, the mean cold gas fraction decreases monotonically 
as $M_*$ grows, and bluer galaxies have higher gas fractions. Our simulated dwarfs appear to follow all these trends. Dopey, in particular, with a stellar mass 
of less than $10^5\,\msun$, has the highest cold gas fraction, $M_{\rm HI}/M_*\simeq 20$. Note that Grumpy has cold gas to stellar mass ratio 
that is similar to that of Leo P, the star-forming extremely metal-poor dwarf recently discovered in the Local Volume by the Arecibo ALFALFA survey \citep{Giovanelli13}. We shall see 
below that Grumpy and Leo P have similar oxygen abundances as well.

\begin{figure*}
\centering
\includegraphics[width=0.495\textwidth]{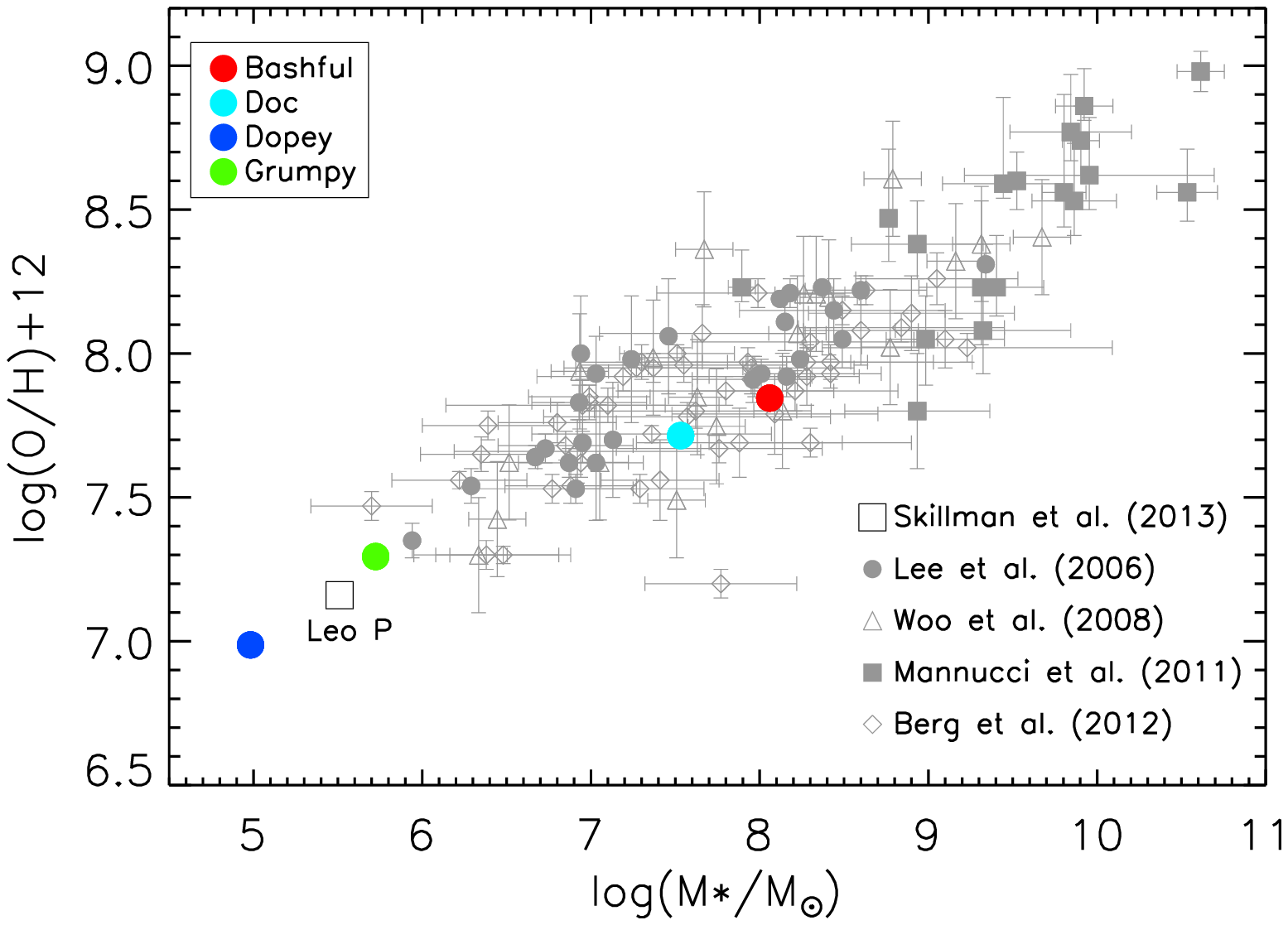}
\includegraphics[width=0.495\textwidth]{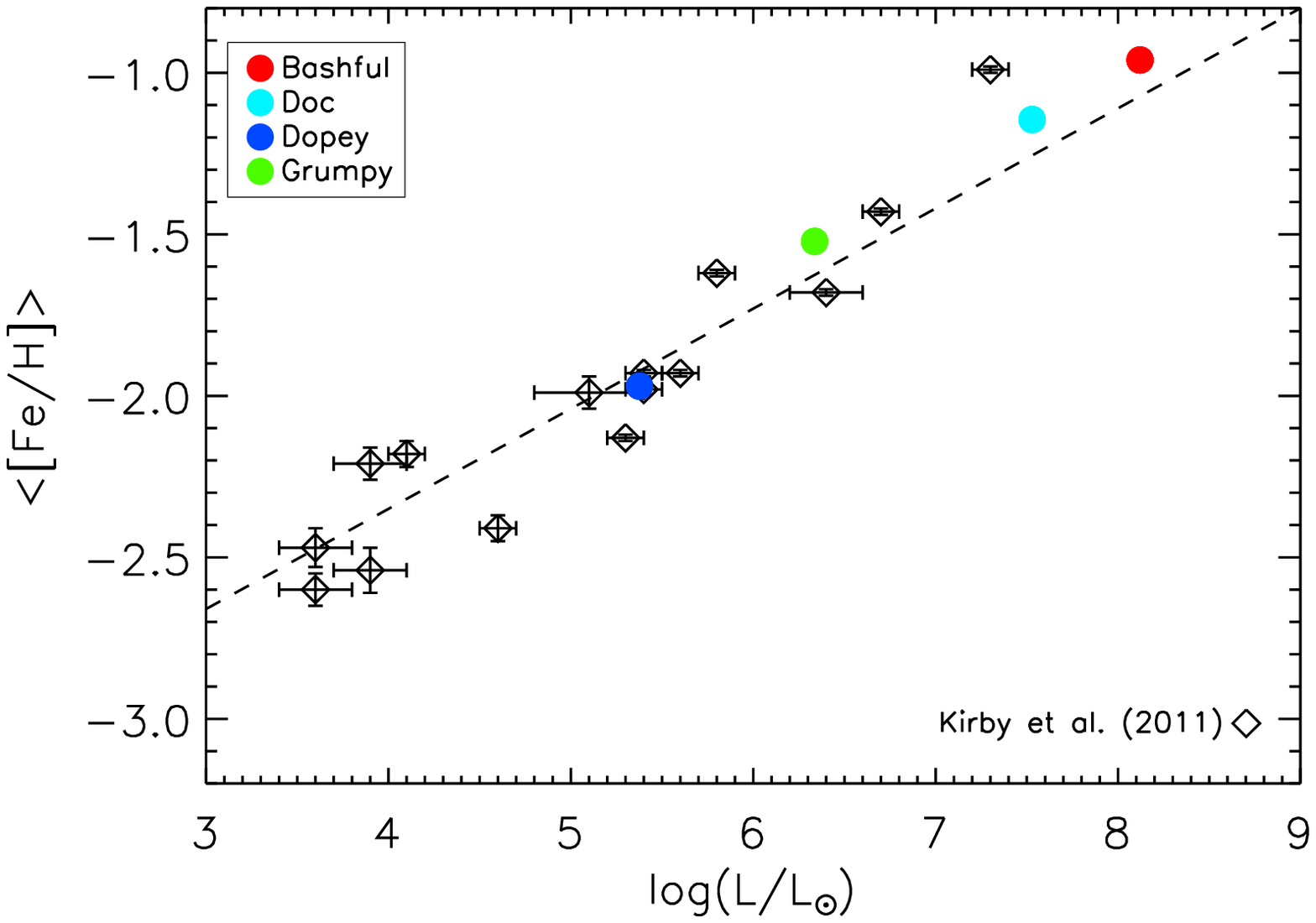}
\caption{{Left panel:} The stellar mass-gas phase metallicity relation of low-mass galaxies. {Filled squares:} host galaxies of $z<1$ long gamma-ray 
bursts \citep{Mannucci11}. {Empty triangles:} Local Group dwarfs \citep{Woo08}. {Filled gray dots:} nearby dwarf irregulars \citep{Lee06}. 
{Empty diamonds:} low-luminosity galaxies in the local volume \citep{Berg12}.   
{Empty large square:} Leo P \citep{Skillman13}.  {Filled colored dots:} simulated dwarfs at $z=0$. All stellar masses have been corrected 
to the same Kroupa IMF. The mean gas metallicity in the simulation is defined by all gas particles with temperatures in the range $5,000-15,000$ K within the 
half-light radius. {Right panel:} The stellar metallicity-$V$-band luminosity relation for Milky Way's dSphs from \citet{Kirby11}. The dashed line is the orthogonal 
regression fit $\langle{\rm [Fe/H]}\rangle=-2.04+0.31\log (L/10^5\,L_\odot)$ of \citet{Kirby11}.
}  
\vspace{+1.0cm}
\label{fig5}
\end{figure*}

We have measured the kinematics of cold gas in our dwarfs for comparison with the observations. In local dIrrs, many of them located in the Local 
Group, \HI\ velocity dispersions in the narrow range $\sigma=6-12\,\kms$ are measured across nearly three orders of magnitude in stellar mass
\citep{Stilp13}. Our simulated galaxies have gas velocity dispersions ranging from 4 to 12 $\kms$, consistent with the observations with the only exception 
of Dopey, whose velocity dispersion is below 2 $\kms$. We note that there is no consensus on what drives the \HI\ 
velocity dispersions in gas-rich dwarfs, and that the physics implemented in our simulations might still
be insufficient to capture ISM turbulence. If supernovae-driven turbulence, one of the postulated mechanisms, is indeed a key factor, then 
it is not surprising that Dopey has an unusually kinematically cold ISM since the amount of star formation present in this system is very modest.
Nevertheless, our simulations reproduce quite well the observed trend of increasing ratio between gas rotational velocity and velocity dispersion, 
$v_{\rm rot}/\sigma$, with increasing luminosity/mass, which
has been documented for more than a decade in nearby dwarfs \citep[see, e.g.,][]{Mateo98}, with the least luminous massive such as Leo A or 
GR8 having $v_{\rm rot}/\sigma$ close to unity \citep{Young96,Carignan90}, similar to Grumpy, and the more massive systems, such as NGC6822 or 
NGC3109, having $v_{\rm rot}/\sigma>2$, such as Bashful and Doc. We note, again, that Dopey has a high $v_{\rm rot}/\sigma > 2$ despite its small mass, 
a result that stems from its unusually low velocity dispersion.

\subsection{Metal Abundances}
\label{metals}
The relationship between the chemical abundances and the stellar content of galaxies is a key diagnostic of their star formation efficiency, infall of 
pristine gas, and outflows of enriched material, and dwarf galaxies are unique laboratories for understanding star formation in nearly pristine low-metallicity 
environments. Observations of galaxies' current gas-phase metallicities typically use oxygen abundances measured in \HII\ regions. Figure \ref{fig5} (left panel) 
shows the gas-phase oxygen abundance versus stellar mass for our dwarfs at $z=0$. The simulated galaxies follow the known 
trend between metallicity and stellar mass (the mass-metallicity relation) observed over 5 dex in stellar mass, from the brightest spirals to the 
dIrrs \citep{Tremonti04,Lee06,Woo08,Berg12}. For a meaningful comparison with the observations (where gas-phase metallicities are measured in 
\HII\ regions), we have only used in the figure gas particles with temperatures in the range $5,000-15,000$ K within the half-light radius of each simulated dwarf. 
Note that, by using solar metallicity yields \citep{Woosley95} in these sub-solar dwarf galaxies, we may overestimate the oxygen abundances, and that measured absolute 
metallicities may be subject to up to a $\Delta$[log(O/H)]=0.7 dex uncertainty depending on the calibration used \citep{Kewley08}.  Our low-mass Dopey and Grumpy dwarfs 
have O abundances, log (O/H)$+12=7.0$ and 7.3, that bracket that of Leo P, one of the lowest metallicity galaxies known \citep{Skillman13}.


Dwarf galaxies are not expected to evolve as ``Closed Box". Although both gas accretion and outflows can change a galaxy's metallicity and gas fraction, 
chemical evolution models show that only gas-rich systems with low star formation rates such as dIrrs can produce and maintain low effective yields \citep{Dalcanton07}.
Indeed, the effective oxygen yields, 
\begin{equation}
y_{\rm O}\equiv {Z_{\rm O}\over \ln (1/f_{\rm gas})},
\end{equation}
of our simulated dwarfs are highly subsolar, with $y_{\rm O}/Z_{\rm O,\odot}=$ 0.10, 0.14, 0.58, 0.08 ($Z_{{\rm O},\odot}=0.0057$, \citealt{Asplund09}) for Bashful, Doc, Dopey, and 
Grumpy, respectively, and correlate with the cold gas-to-stellar mass ratio. They drop well below the Closed Box ``true" value of $2.4\,Z_{{\rm O},\odot}$ because of 
changes both in the gas-phase mass abundance of oxygen, $Z_{\rm O}$, and in the gas mass fraction $f_{\rm gas}=M_{\rm gas}/(M_{\rm gas}+M_*)$, where 
$M_{\rm gas}=1.36 M_{\rm HI}$ is the total cold gas mass including helium. While these values are comparable to the range of effective yields for dwarf galaxies 
reported by \citet{Lee06}, we notice here that strong metal-enriched galactic outflows are not ubiquitous in the simulated dwarf galaxies: the fraction of all the metals 
ever produced that is {\it retained} at the present-epoch by our dwarfs, $f_Z$, {\it increases with decreasing stellar mass}, with $f_Z=$10, 11, 92, and 46 percent for 
Bashful, Doc, Dopey, and Grumpy, respectively. With the lowest metallicity and star formation efficiency, a peak star formation 
rates not exceeding $0.004\,\mdot$ (100 times lower than in Bashful), and the highest gas-to-stellar mass ratio, Dopey is simply unable to power 
strong galactic-scale outflows.  

Although star formation in Dopey and Grumpy is linked to interactions, both are not star-forming today nor exhibit strong gas inflows (Figure \ref{figtidal}). Thus, their low chemical abundances
are mainly the result of inefficient star formation rather than metal enriched-outflows or the infall of pristine gas. \citet{Skillman13} suggest that there are two classes of extremely metal-deficient 
galaxies (XMDs): 1) the more quiescent XMDs such as Leo P, in which the low metallicity is due to their small halo mass; and 2) the rarer starburst galaxies like I Zw 18, whose low metallicity is 
probably associated with pristine gas inflows \citep{Ekta10} and which fall below the mass-metallicity relation. Dopey and Grumpy are closer to the former, more quiescent XMDs at the present epoch. 
Their star formation histories and structural properties, such as the very low central surface brightnesses, make them akin to the faint Local Group dIrrs such as Leo A (M$_{B} \sim$ -11), 
which have typically $\mu{_B} > 23 \  {\rm mag~arcsec^{-2}}$ \citep{Mateo98}. 

\begin{figure}
\centering
\includegraphics[width=0.495\textwidth]{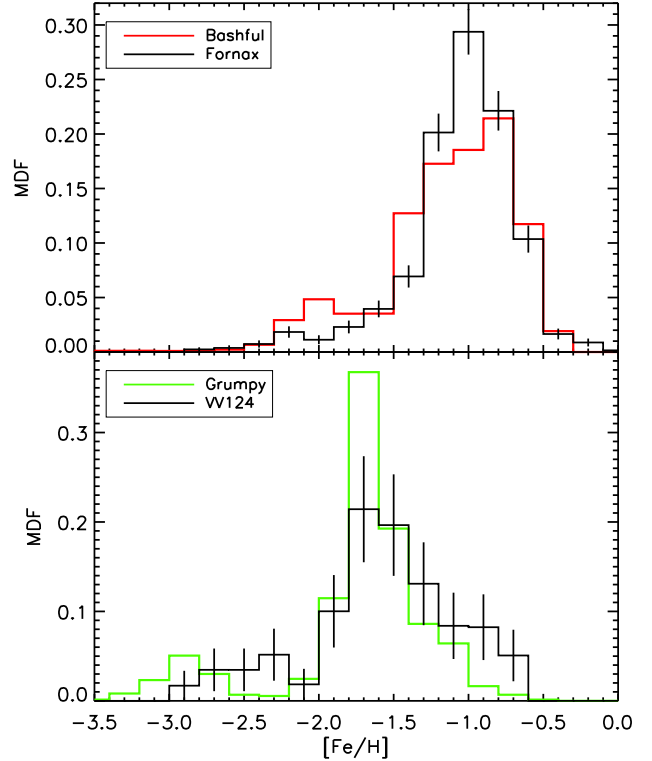}
\caption{The differential stellar metallicity distribution function (MDF) of Bashful ({top}) and Grumpy ({bottom}) compared to the MDFs of Fornax \citep{Kirby11}
and VV124 \citep{Kirby12}. 
}  
\vspace{+0.cm}
\label{figMDF}
\end{figure}

While the gas-phase metallicity traces the metallicity of the latest generation of stars to form, the (lower) stellar metallicity represents an average over the entire
star formation history of a galaxy. The right panel of Figure \ref{fig5} shows that, despite being gas-rich, the average stellar metallicity of our simulated dwarfs 
follows remarkably well the $V$-band luminosity-metallicity relation defined by the Milky Way's dSph satellites \cite{Kirby11}. This suggests that chemical enrichment 
proceeds largely independent of environment and is primarily dictated by the mass of the galaxy. The idea that dSphs and dIrrs of the same luminosity 
may have similar global metal abundances is consistent with the recent observations of the isolated dIrrs VV124 \citep{Kirby12} and WLM \citep{Leaman13}, which  
are found to lie on the Local Group luminosity-metallicity relation. The differential stellar metallicity distribution functions (MDFs) for Bashful and Grumpy are 
shown in Figure \ref{figMDF}. The MDFs relate more directly to the chemical enrichment histories of dwarfs than their bulk metallicity properties, and can be dependent on the 
assumed IMF and the implementation of feedback and metal mixing in the ISM \citep{Pilkington12}.
For illustrative purposes only, we have compared the simulated distributions with the observed MDFs of a dSph, Fornax \citep{Kirby11}, and 
a dIrr, VV124 \citep{Kirby12}. These two systems have comparable mean metallicities to Bashful (Fornax) and Grumpy (VV124) (but different luminosities and SFHs), and 
as it is typical for large dSphs, their metallicity distribution is known to fit a chemical evolution model with infalling gas better than a Closed Box model 
\citep{Kirby11,Kirby12}. While Bashful has a slightly wider metallicity peak than Fornax, and Grumpy has a more prominent extremely metal-poor, 
[Fe/H]$<-3$, tail compared to VV124, overall the simulated dwarfs appear to have realistic metallicity spreads and distributions. 
Extremely metal-poor stars have been observed in Milky Way's ultra-faint dwarf in classical dSphs \citep[e.g.,][]{Kirby08,Kirby09}.   
We defer a more quantitative comparison between observed and theoretical MDFs to a future paper that will also make use of a larger sample of simulated dwarfs, and explore the 
sensitivity of our results to changes in the IMF, feedback, and metal mixing schemes.

We finally note that, while the baryon fractions of the four dwarfs discussed above are all within a factor of two of each other, their star formation efficiencies and 
cold gas fractions {\it span two orders of magnitudes}. Therefore, as already noted by \citet{Munshi13}, low star formation efficiencies 
are not simply the result of the {\it blowing away} of all the baryons from the host potential wells. Baryons are retained but are unable to make stars because
of the more realistic description of where stars form (in high density clouds) and how feedback regulates the thermodynamics of the ISM without leading to an 
excessive outflow mass loading factor.

\subsection{Three Dark Dwarfs}

The three lightest dwarfs (``Happy", ``Sleepy", and ``Sneezy", all with $M_{\rm vir}<10^9\,\msun$), do not form any stars 
down to our stellar mass resolution limit of $10^3\,\msun$, i.e.  their stellar mass fractions fall below $\lta 2\times 10^{-6}$. They are also baryon poor, 
with baryon fractions that are well below 0.5\% today (partly as a consequence of gas removal by ram-pressure stripping), and never exceeded 3\% in the past. 
Throughout cosmic history, their central gas column densities (simply estimated by multiplying 
the maximum gas volume density reached at a given redshift by the SPH smoothing length) remain below $0.05\,\msun$ pc$^{-2}$, far too low to self-shield against the UVB and
build up a substantial \HI\ mass. The twin simulation run without turning the UVB on shows that the fate of such systems is indeed regulated by external photoheating.
We find that, without photoionization inhibiting the condensation process of gas in galaxy halos, Happy, Sleepy, and Sneezy would all be forming stars but only at $z\lta 2$, 
and with very low efficiencies. We note here that these three dwarfs appear to be the low-mass low-redshift counterparts of the population of 
gas-rich ``dark galaxies" recently studied by \citet{Kuhlen13} and detected in a deep narrow-band survey for Ly$\alpha$ emission by \citet{Cantalupo12}. 
Using a cosmological hydrodynamic simulation with the \textsc{Enzo} code of the formation of dwarf galaxies 
at redshifts $z\gta 2.5$, \citet{Kuhlen13} have shown that a star formation prescription regulated by the local H$_2$ abundance leads to the suppression of star formation in dwarf
galaxy halos with $M_{\rm vir}<10^{10}\,\msun$. \citet{Kuhlen13} dark galaxies form late and their gaseous disks never reach the surface densities, $\gta 750\,\msun\,
{\rm pc}^{-2}\, (Z/10^{-2}\Zsun)^{\!-0.88}$, that are required to build a substantial molecular fraction. Analogously, in our simulation with a star 
formation prescription based on total gas density, the ISM of Happy, Sleepy, and Sneezy remains always above $10^{4}$ K and never condenses to gas densities 
in excess of $10^{-2}\,$cm$^{-3}$. 

\subsection{Circumgalactic Medium}

\begin{figure*}
\centering
\includegraphics[width=0.95\textwidth]{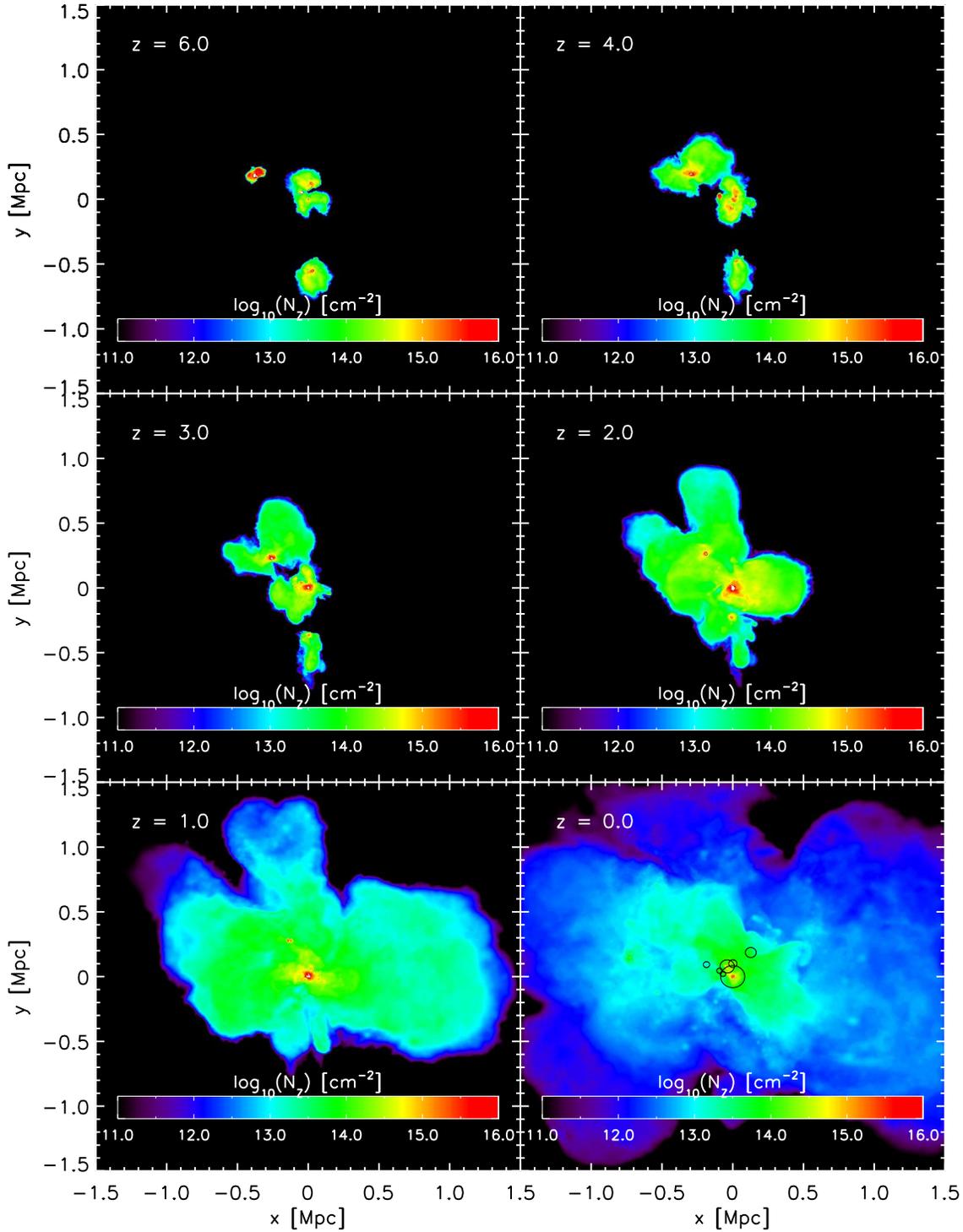}
\vspace{-0.5cm}
\caption{The growth of the metal-enriched bubble around our simulated dwarfs. The figure shows total metal column density (in units ofcm$^{-2}$) at different redshifts in a 
cube of 3 comoving Mpc on a side.  At $z=0$, the virial radii of our seven dwarfs are marked by the black circles. By $z=0$, $>87$\% of all the metals produced are spread 
over a region of 3 comoving Mpc across, beyond the dwarfs' CGM and out into the IGM.}  
\label{figCGM}
\end{figure*}

Simulations of the ionization, chemical, thermodynamic, and kinematic state of gaseous material in the CGM of dwarfs hold clues to understanding the exchange of mass, metals, and 
energy between the smallest, most abundant, and less luminous galaxies and their surroundings. It is the flows of gas into such galaxies, and from them back into their environments, 
that determine the response of baryons to their shallow potential wells and enrich intergalactic gas. While a detailed study of the CGM of dwarfs and of the baryonic outflows and 
inflows that shape their central dark matter density profiles is postponed to a subsequent paper of this series, it is interesting at this stage to briefly comment on the chemical 
imprint left on the environment by our dwarfs. This is clearly illustrated in Figure \ref{figCGM}, where we show the growth of a large metal-enriched bubble in the simulated 
galaxy forming region. The figure shows projected gas particle metallicity at different redshifts in a cube of 3 comoving Mpc on a side. A total mass of $5.7 \times 10^{6}\,\msun$ of 
heavy elements, more than 87\% of those produced over cosmic history, has been spread over such region, well beyond the CGM of the dwarfs and out into the IGM. 
While the majority of stars form at $z>1$ in Bashful and Doc (the main contributors of heavy elements), the metal-enriched bubble continues to grow in comoving coordinates down 
to the present time, as the ejected wind material is unbound and not recycled back into the galaxies.  A low gas recycling fraction in dwarf galaxies is also found by 
\citet{Brook13} in the MaGICC simulations that use the same blastwave scheme for supernova feedback.

The size of the metal bubble (defined here as the impact parameter from the center of Bashful where the covering fraction of material with total metal column density $N_Z > 10^{12}\,\cmm$ 
decreases below 50\%) remains around 6-7 Bashful's virial radii ($R_{\rm vir,B}$) at  $z\gta 2$, increases to 10 $R_{\rm vir,B}$ at $z\approx 1$, and eventually reaches 16 
$R_{\rm vir,B}$  (1.4 Mpc) at redshift 0. For comparison, outflows in the late-type massive spiral ``Eris2", simulated with the same code and similar prescriptions \citep{Shen13}, 
produce a metal bubble extending from four host virial radii at $z\gta 1$ to 6 virial radii at $z\lta 0.5$. The mass loading factor at the virial radius, $\dot M_w/\dot M_*$,
(characterizing the rate at which material is ejected per unit star formation rate), can be as high as 12 for Bashful and 17 for Doc, more than one dex higher than in Eris2  and 
comparable to the mass loading of Eris2's satellites \citep{Shen12}. Clearly, although less efficient at forming stars than more massive systems, {\it dwarf galaxies appear to be 
collectively more effective at enriching the low density IGM on large scales.} 

\begin{figure*}
\centering
\includegraphics[width=0.80\textwidth]{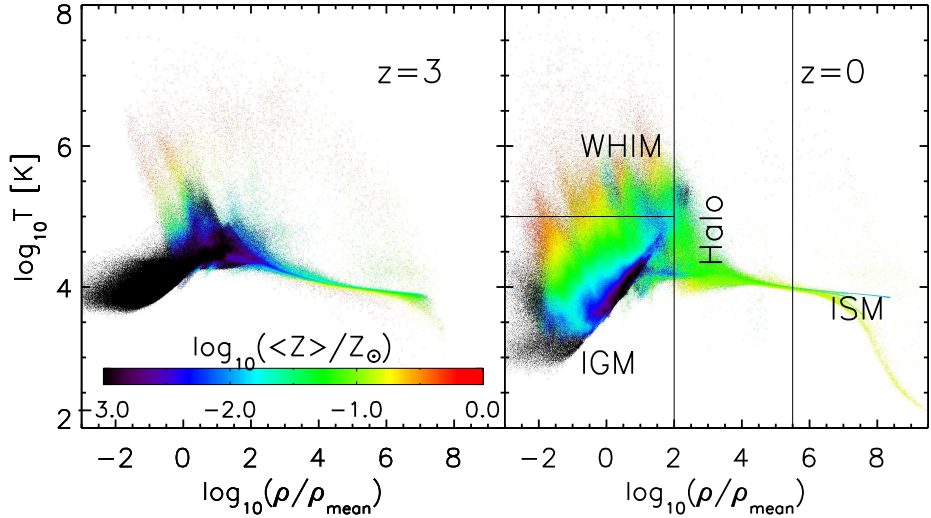}
\vspace{0.8cm}
\caption{Distribution of all enriched gas in the temperature-density plane at $z=3$ (left panel) and $z = 0$ (right panel) within the simulated volume. The color scale 
indicates mass-weighted metallicity. In the right panel, we define the interstellar medium as gas with overdensity $\delta \equiv \rho/\rho_{\rm mean} > 10^{5.5}$, corresponding 
to $n>0.1$ atoms cm$^{-3}$, and halo gas as having $100<\delta<10^{5.5}$. All gas with $\delta<100$ belongs to the IGM, and we use the term WHIM to define IGM gas with $T>10^5$ K. }  
\vspace{+0.cm}
\label{figphase}
\end{figure*}

The mass-weighted distribution of all enriched gas within the simulated region in the temperature-density plane is shown in Figure \ref{figphase}. Heavy elements are spread 
over a large range of phases, from cold star-forming material at $T<10^4$ K and $n>100$ atoms cm$^{-3}$ (corresponding to an overdensity $\delta=\rho/\rho_{\rm mean}>6\times 10^6$ at $z=3$ and 
$\delta>4\times 10^8$ at $z=0$) to hot, $T>10^6$ K low density intergalactic gas that cannot cool radiatively over a Hubble time. The black strip in the lower left corner of the 
figure marks the nearly pristine, adiabatically cooling IGM, while the colored swath in the lower right corner marks metal-rich gas in the dense inner regions of the dwarfs. 
Hot gas is vented out in the CGM by the cumulative effect of SN explosions, expands first into a warm-hot intergalactic medium (WHIM, $T>10^5$ K), and propagates further out
in intergalactic space, cooling down and enriching the cool IGM with heavy elements.  
Although the WHIM has the highest metallicity, the majority of heavy elements by mass are located in the cool IGM component. 
Specifically, the fractional mass of gas-phase heavy elements locked in the cold ($T<3\times 10^4$ K), warm ($3\times 10^4 {\rm K}<T<3\times 10^5$ K), and 
hot ($T>3\times 10^5$ K) phases is 57\%, 23\% and 20\%, respectively.  This is in contrast with the CGM around more massive star-forming galaxies, 
where more than half of the metals are locked in the hot component instead \citep{Shen13}.  
More than 55\% of all gas-phase metals produced lie outside the virial radii of the 4 luminous dwarfs at redshift three, increasing to $>87\%$ by the present epoch. 
The majority of the metals in all three phases reside in underdense regions with $\delta<1$ at $z=0$. As has been done for more massive halos \citep{Stinson12,Shen13}, 
a more detailed comparisons with observations of gas and metal distributions around dwarf galaxies will be presented in a future study. 

\section{Summary}

We have presented results from a fully cosmological, very high-resolution, $\Lambda$CDM ``zoom-in" simulation of a group of seven field dwarf galaxies
with present-day virial masses in the range $4.4\times 10^8-3.6\times 10^{10}\,\msun$. The simulation includes a blastwave scheme for supernova feedback, a star
formation recipe based on a high gas density threshold, metal-dependent radiative cooling, a scheme for the turbulent diffusion of metals and thermal energy, and a 
uniform UV background that modifies the ionization and excitation state of the gas. The properties of the simulated dwarfs appear strongly modulated by the depth of the
gravitational potential well. All three halos with $M_{\rm vir}<10^9\,\msun$ are devoid of stars, as they never reach the density threshold for star formation
of 100 atoms cm$^{-3}$, their fate being regulated by external photoheating. These three dwarfs appear to be the low-mass low-redshift counterparts of the population of 
gas-rich ``dark galaxies" recently studied by \citet{Kuhlen13} and discovered at high redshift by \citet{Cantalupo12}. The other four, $M_{\rm vir}>10^9\,\msun$ dwarfs have 
blue colors, low star formation efficiencies, high cold gas to stellar mass ratios, and low stellar metallicities. While the baryon fractions of the four dwarfs discussed above 
are all within a factor of two of each other, their star formation efficiencies and cold gas fractions span two orders of magnitudes. Low star formation efficiencies are not simply the result of the 
blowing away of all the baryons from the host potential wells. Baryons are retained but are unable to make stars because of the more realistic description of where stars 
form (in high density clouds) and how feedback regulates the thermodynamics of the ISM without leading to an excessive outflow mass loading factor.

The star formation histories of our dwarfs are very bursty, characterized by peak specific star formation rates in excess of $50-100$ Gyr$^{-1}$, far outside
the realm of normal, more massive galaxies, and in agreement with observations of extreme emission-line starbursting dwarfs by the CANDELS survey \citep{vanderWel11}. 
The median stellar age of the simulated galaxies decreases with decreasing halo mass, with the two $M_{\rm vir}\simeq 2-3\times
10^{9}\,\msun$ dwarfs being predominantly young, hosting stellar populations of age a few Gyrs, and the two more massive systems hosting substantial amounts of intermediate age and old 
stars.  The two cosmologically young dwarfs are lit up by tidal interactions, and have compact morphologies and colors similar to those of XBCDs. They have low surface brightnesses, however, 
and are not undergoing starbursts nor exhibit strong gas inflows at the present epoch, thus may not be detectable through emission line surveys with current detection limits \citet{Papaderos08}. 
Many of their properties, as well as their late star formation histories, resemble those of the local dIrr Leo A \citep{Cole07}. Moreover, their metallicity does not fall below the mass-metallicity 
relationship as in the starburst XBCD I Zw 18, but are reminiscent of Leo P, the relatively quiescent star-forming extremely metal-poor dwarf recently discovered
in the Local Volume by the Arecibo ALFALFA survey \citep{Skillman13}. This suggests a strong correlation between gas inflows, metallicities, and star formation rates. 
One of the two faint galaxies (``Grumpy") has also a similar cold gas-to-stellar mass ratio to that of Leo P \citep{Giovanelli13}.

The simulated dwarfs lie on the stellar mass-halo mass relation of \citet{Behroozi13}, the cold gas-stellar mass relation of \citet{Huang12}, the 
stellar mass-gas phase metallicity relation of low-mass galaxies of \citet{Woo08} and \citet{Lee06}, and the stellar metallicity-$V$-band luminosity relation 
for Milky Way's dSphs of \citet{Kirby11}. They have realistic metallicity spreads and distributions, cumulative star formation histories in broad agreement  
with those of dIrrs in the ANGST sample \citep{Weisz11}, and a dynamically hot stellar structure as recently observed in the isolated dIrr galaxy WLM \citep{Leaman12}.

Metal-enriched galactic outflows produce highly sub-solar effective yields but are not sufficient to completely quench star formation activity and
are not ubiquitous: the fraction of all the metals ever produced that is retained at the present-epoch by our dwarfs increases with decreasing stellar mass.
``Dopey", the faintest dwarf with the lowest metallicity and star formation efficiency, a peak star formation rates not exceeding $0.004\,\mdot$, and the highest 
gas-to-stellar mass ratio, is simply unable to power strong galactic-scale outflows. By $z=0$, more than 87\% of all the metals produced by our most luminous dwarfs 
over cosmic time are spread over a region of 3 Mpc across centered on the most luminous system, well beyond the CGM of our dwarfs and out into the IGM. Although 
less efficient at forming stars than more massive systems, this suggests that dwarf galaxies may be collectively most effective at enriching the low density IGM on large scale. 

While our calculations appear to reproduce quantitatively at least some of the complex baryonic processes that regulate the ``metabolism" of dwarf galaxies, 
the small size of our sample, the lack of explicit radiation transport, and the simplistic feedback scheme we adopt must all be acknowledged as limitations to the work 
presented here. A detailed study (and comparison with the observations) of the stellar and gas kinematics and central dark matter density profiles of a larger
sample of simulated dwarfs is required to fully validate our 
approach, and has been postponed to a subsequent paper of this series. In spite of such caveats, we find it encouraging that cosmological simulations of the 
formation of dwarf galaxies in $\Lambda$CDM appear to simultaneously reproduce the observed stellar mass and cold gas content, resolved star formation 
histories, stellar kinematics, and metallicities of field dwarfs in the Local Volume.
 
\acknowledgments
We thank the anonymous referee for his/her though-provoking questions and useful suggestions.  We thank D. Weisz for providing the ANGST data plotted in Figure 3. Support for this work 
was provided by the NSF through grants OIA-1124453 and AST-1229745, and by NASA through grant NNX12AF87G (P.M.). F.G. was supported by NSF through grants AST-0908499 and AST-1108885. C.C. 
acknowledges support from the Alfred P. Sloan Foundation.

\end{document}